\documentclass[pre,aps,preprint,showpacs]{revtex4}
\usepackage{revsymb,latexsym,amssymb,amsmath,comment}
\usepackage{graphicx,psfrag,subfigure,stmaryrd}

\newcommand{\be}{\begin{equation}}
\newcommand{\ee}{\end{equation}}
\newcommand{\bel}[1]{\begin{equation}\label{#1}}
\newcommand{\bea}{\begin{eqnarray}}
\newcommand{\eea}{\end{eqnarray}}
\newcommand{\ba}{\begin{array}}
\newcommand{\ea}{\end{array}}
\newcommand{\bef}{\begin{figure}}
\newcommand{\ef}{\end{figure}}

\begin{document}

\author{Thomas Bose and Steffen Trimper}
\affiliation{Institute of Physics,
Martin-Luther-University, D-06099 Halle Germany}
\email{thomas.bose@student.uni-halle.de}
\email{steffen.trimper@physik.uni-halle.de}
\title{A stochastic model for tumor growth with immunization}
\date{\today }
\begin{abstract}
We analyze a stochastic model for tumor cell growth with both multiplicative and additive colored noise as well as a 
non-zero cross-correlations in between. Whereas the death rate within the logistic model is altered by a deterministic 
term characterizing immunization, the birth rate is assumed to be stochastically changed due to biological motivated 
growth processes leading to a multiplicative internal noise. Moreover, the system is subjected to an external additive 
noise which mimics the influence of the environment of the tumor. The stationary probability distribution $P_s$ is derived 
depending on the finite correlation time, the immunization rate and the strength of the cross-correlation. $P_s$ offers a 
maximum which becomes more pronounced for increasing immunization rate. The mean-first-passage time is also calculated in 
order to find out under which conditions the tumor can suffer extinction. Its characteristics is again controlled by the 
degree of immunization and the strength of the cross-correlation. The behavior observed can be interpreted in terms of 
the three state model of a tumor population.

\end{abstract}

\maketitle

\section{Introduction}

A fundamental aspect of all biological systems is the understanding of emergence of  
cooperative behavior. The competitive interaction among different growth and death processes 
and the inclusion of external mechanism are widely believed to influence the global properties of 
such systems \cite{murray}. In this context much effort has been devoted to model the dynamics of 
competing population through a nonlinear set of rate equations such as proposed by Lotka and Volterra
\cite{lo,vo} or a broad variety of their variants as a stochastic model for ecosystems \cite{cl}, 
coexistence versus extinction \cite{rmf} or special clustering in Lotka-Volterra model \cite{plg}. 
Prey-predator systems are likewise related to that kind of models, where recently also fluctuations 
and correlations are discussed  \cite{mgt,rs} as well as instabilities with respect to spatial distributions \cite{as}. 
The heuristic approach is based upon deterministic evolution equations. Otherwise, a population of proliferating 
cells is a stochastic dynamical system far from equilibrium \cite{bs}. Proteins and other molecules are produced and degraded 
permanently. Cells grow, divide and inherit their properties simultaneously to the next generation.  
To gain some more insight into the generic behavior of phenomena such as tumor cell growth, it is desirable 
to take into account both internal and external stochastic noises as well as spatial correlations.
 
In the present paper we are interested in tumor growth which had been attracted attention over several decades. 
Mathematical modeling of the growth of a certain population is based on different equations where the logistic growth and the 
Gompertz law are the most popular deterministic models \cite{mbf}. A more refined model was presented in  
\cite{mxz}, however we argue that the solutions for the stationary probability distribution and the mean-first-passage 
time are not calculated correctly. The details and the corrections are given in our paper in Sec.III and IV. Nevertheless, 
the model in \cite{mxz} includes already both additive and multiplicative noise terms considered likewise in \cite{qw}. 
However, the stationary distribution function presented in that paper is also not correct as pointed out in \cite{bfr} 
and replied in \cite{qw2}, see also our results discussed below. The role of pure multiplicative noise may induce stochastic 
resonance, which appears in an anti-tumor system \cite{zsh}. In that work the deterministic forces are modified 
as it will be also discussed in the present paper. The mean first passage time of a tumor cell growth is altered by cross correlations 
of the noise, see already \cite{wwm}. Essentially for tumor modeling is also the inclusion of therapy elements as proposed 
in \cite{kbbd}. In our model we analyze a special immunization term which enhances the the death rate. Another 
model \cite{zslwh} is devoted to the spatiotemporal triggering infiltrating tumor growth. 

Our approach can be grouped into the permanent interest in a statistical modeling of growth model, where evolution equations 
of Langevin or Fokker-Planck-type play an decisive role \cite{gardiner}. In particular, the focus is concentrated on 
correlated colored noises \cite{wlk} in the form of multiplicative noise \cite{jl} and additive noise \cite{z}. 
A similar approach is also applied in \cite{cb} for the Bernoulli-Malthus-Verhulst model. In the 
context of population dynamics different aspects has been studied such as time delay effects \cite{nm}, a general 
classification scheme for phenomenological universality in growth problems \cite{cdg}, extinction in birth-death-systems \cite{am}, 
the complex population dynamics as competition between multiple-time-scale phenomena \cite{bdsp} and the the dissipative branching 
in population dynamics \cite{jms}. 

The goal of our paper is inspired with the view to alter the models in such a manner that both immunization and correlated noise are 
included. Especially, we want to demonstrate that a finite correlation time and a nonzero immunization rate have an significant 
impact on the different steady states realized within the model. Additionally we analyze the interplay between an internal noise 
leading to a stochastic birth rate and an external noise. Furthermore, the mean-first passage time is calculated which enables 
us to analyze under which conditions, depending on the correlation time and the immunization rate, the tumor population can suffer extinction. 
Our paper is organized as follows: In Sec. II we define the Langevin equation with different multiplicative noises and their cross-correlation functions, 
the meaning of that is considered in detail. Then we introduce an immunization term the influence of which will be analyzed in the paper. 
Such an additional term leads to a significantly modified death rate. Based on the related Fokker-Planck equation the stationary probability 
distribution (SPD) is studied in Sec. III. The expression for the mean-first-passage time (MFPT) and its meaning is explained in Sec. IV. 
Further we discuss the relation of our results to real tumor growth. In Sec. V we finish with some conclusions.

\section{The Tumor Model}

In order to develop a statistical tumor growth model, we consider the general type of Langevin equation, that reads
\be	\frac{dx}{dt} = f(x)+g_1(x)\epsilon _1(t)+g_2(x)\epsilon _2(t),
\label{lg}
\ee 
where $x(t)$ denotes the number of tumor cells at time $t$, $f(x), g_1(x)$ and $g_2(x)$ are deterministic functions and 
$\epsilon _1(t)$ and $\epsilon _2(t)$ are colored noises with zero mean and colored cross-correlation. 
These statistical properties are given by $ \langle \epsilon _1(t)\rangle=0;\quad \langle \epsilon _2(t)\rangle = 0$ and the 
corresponding correlation functions
\bea
C(t-t')=\left(\begin{array}{cc}
		\langle \epsilon _1(t)\epsilon _1(t')\rangle & \langle \epsilon _1(t)\epsilon _2(t')\rangle \\
		\langle \epsilon _2(t)\epsilon _1(t')\rangle & \langle \epsilon _2(t)\epsilon _2(t')\rangle
	\end{array}\right)=\left(\begin{array}{cc}
		\frac{M}{\tau _1}\,exp\left(-\frac{|t-t'|}{\tau _1}\right) & \frac{\lambda \sqrt{M\alpha }}{\tau _3}\,exp\left(-\frac{|t-t'|}{\tau _3}\right) \\
		\frac{\lambda \sqrt{M\alpha }}{\tau _3}\,exp\left(-\frac{|t-t'|}{\tau _3}\right) & \frac{\alpha }{\tau _2}\,exp\left(-\frac{|t-t'|}{\tau _2}\right)
	\end{array}\right)\,. 
\label{lg1}
\eea
Here, the elements of the correlation matrix $C_{ij} (t-t')$ are assumed to be symmetric $C_{ij} = C_{ji}$. The quantities $M$ and $\alpha $ are the noise intensities and $\tau _1$ and $\tau _2$ are the correlation times of the autocorrelation functions $C_{11}$ and $C_{22}$. The parameters $\lambda$ and 
$\tau _3$ characterize the strength of the cross-correlation function between $\epsilon _1(t)$ and $\epsilon _2(t)$ and the cross-correlation 
time, respectively. In our model we consider a modified logistic growth model with
\be
f(x) = a x - b(x) x^2 ,\quad b(x) = b_0 + \phi(x) \equiv b_0 + \frac{\beta}{1 + x^2}  
\label{f}
\ee
Here, the parameter $a$ is the deterministic growth rate and $b_0$ denotes the decay rate proportional to the inverses carrying capacity, respectively. 
This death rate is altered by inclusion of a tumor-immunization interaction represented by the function $\phi (x)$ \cite{murray}, where the  
parameter $\beta $ designates the strength of the immunization. Under immunization the effective death rate $b(x)$ is enhanced where the decay of the 
rate depends on the immunization strength $\beta$. The behavior of the effective death rate is depicted in Fig.~\ref{deathrate}.
\bef
\includegraphics[width=0.6\textwidth]{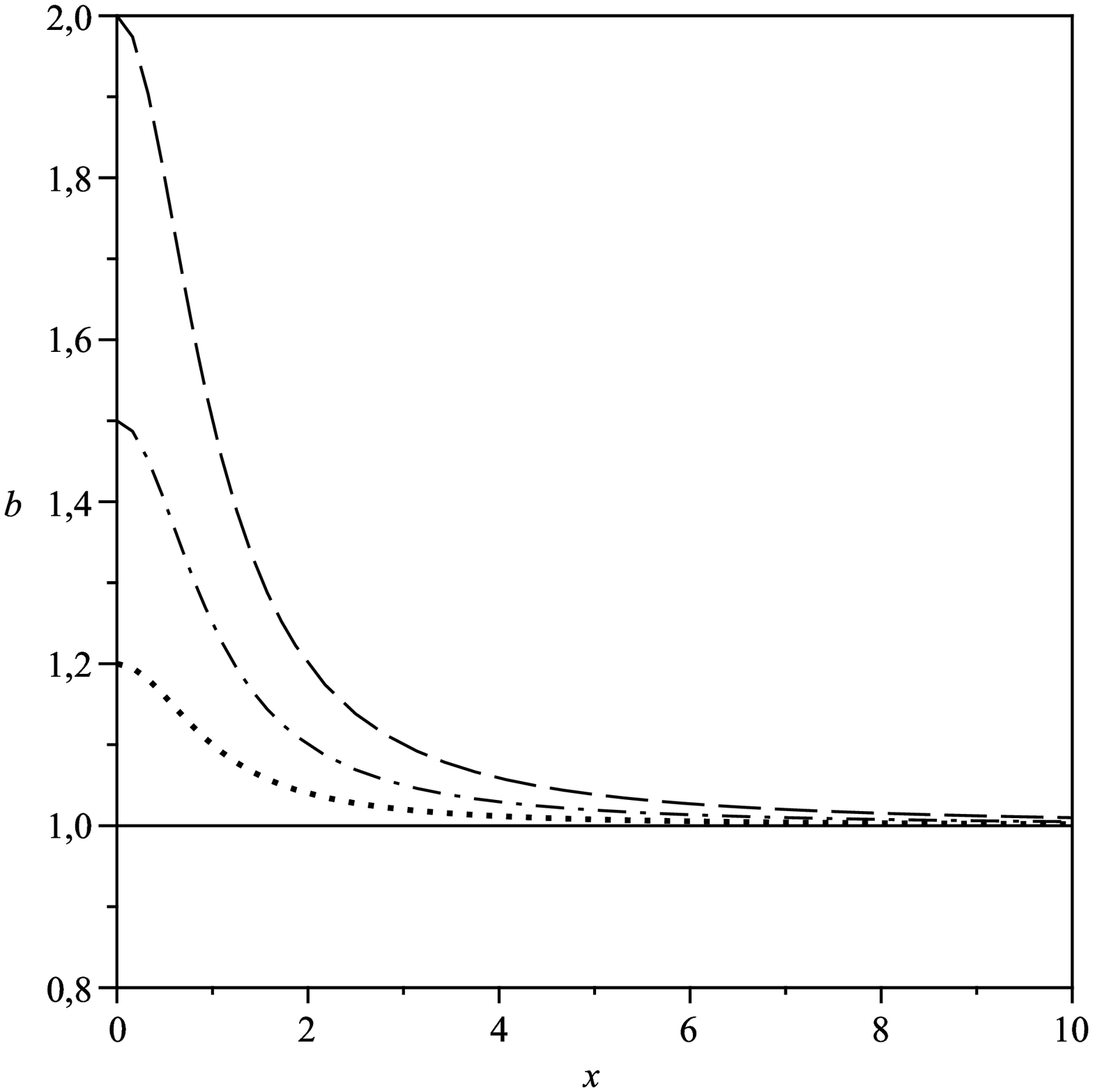}
	\caption{Plot of the death rate $b$ as a function of the cell number $x$ when $b_0=1$ is fixed: $\beta$ takes $0$ (solid line), $0.2$ (dotted line), $0.5$ (dashdotted line) and $1$ (dashed line).}
	\label{deathrate}
\ef
The tumor cell evolution is further coupled to internal and external noises denoted by $\epsilon _1(t) $ and $\epsilon _2(t)$, respectively. 
Whereas the death rate is  systematically enhanced by immunization, modeled by the deterministic function $\phi(x)$, the effective birth rate should be  
influenced by the stochastic force $\epsilon_1(t)$. This leads to the assumption 
\be
g_1(x) = -x\,.
\label{g}
\ee
Furthermore, the effect of additive noise represented by $\epsilon_2(t)$ is incorporated into the system by
\be
g_2(x) = 1\,.
\label{g2}
\ee
Notice that all parameters are dimensionless, so that the prefactors in the last equations could be set as unity.  
With regard to the discussions in Sec.IV let us introduce an effective potential $V(x)$ according to the deterministic force $f(x)$, that reads 
\be
V(x)=-\int{f(x)\, d(x)}.
\label{potV}
\ee
Evaluating (\ref{potV}) yields the following expression for $V(x)$ :
\be
V(x)=\frac{1}{3}b_0 \,x^3-\frac{1}{2}a \,x^2+\beta \,( x- \arctan x ).
\label{Vsol}
\ee
The potential $V(x)$ is presented in Fig.~\ref{GPotV}.
\bef
	\includegraphics[width=0.6\textwidth]{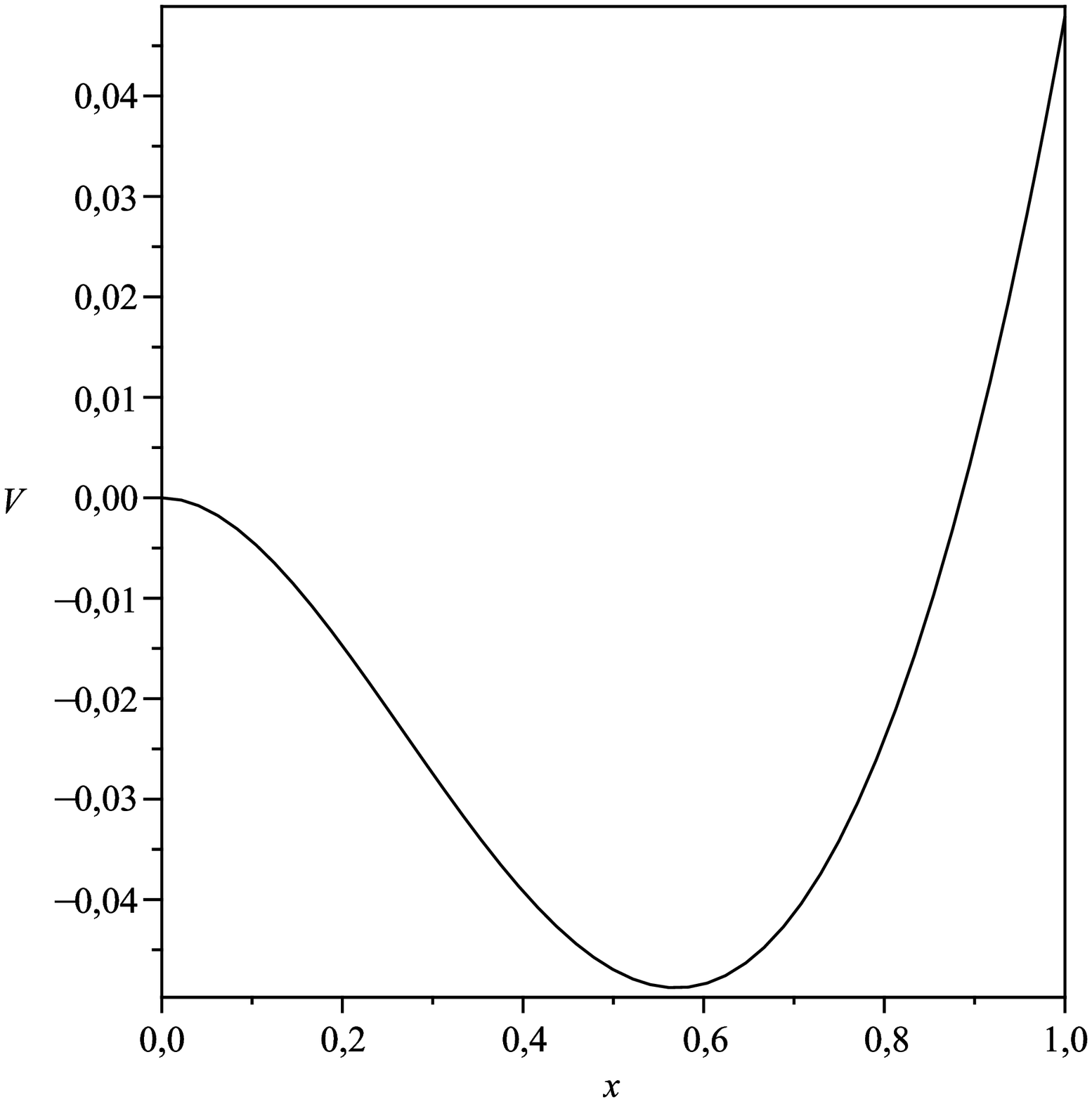}
	\caption{Plot of the effective potential $V$ as a function of the cell number $x$ where $a=1$, $b_0=1$ and $\beta=1$.}
	\label{GPotV}
\ef
The stationary points can be determined by setting $f(x) = - \frac{d}{dx}V(x)=0$. From here we discriminate between four extrema, from  
which only two of them are real in the parameter range considered. The remaining stationary points take complex values and will not be discussed furthermore.
Thus, we have derived a potential with a minimum at $x_1 = x_s > 0$ and a maximum at $x_2 = 0$. 

\section{Fokker-Planck-Equation }

\subsection{Derivation of the stationary probability distribution (SPD)}

As a next step the Langevin Equation (\ref{lg}) is transformed into an equivalent Fokker-Planck equation (FPE) \cite{vk,gardiner,mxz,z}. 
To that aim let us consider $x(t)$ as a random variable whose probability density function $\rho(w,t)$ is a delta-function 
$$
\rho(w,t) = \delta( x(t) - w)\,.
$$ 
From here one can find the stochastic Liouville equation \cite{vk} for the probability distribution function
\be
P(w,t)=\langle \rho (w,t) \rangle\,.
\label{d}
\ee
Here, $P(w,t)$ is the density of the probability distribution function that the process $x(t)$ takes the value $w$ at time $t$. 
From this relation combined with Eqs.~\eqref{lg} - \eqref{g2}, one obtains the FPE in the form
\bea
\frac{\partial P(w,t)}{\partial t}=-\frac{\partial }{\partial w}[A(w)P(w,t)]+\frac{\partial ^2}{\partial w^2}[B(w)P(w,t)].
\label{afpe}
\eea
The explicit expressions for $A(w)$ and $B(w)$ are
\bea
A(w)&=& \left( \, a + \frac{M}{1+a\tau _1 }\right) \,w - \left( \, b_0  + \beta \frac{1}{1+w^2} \right) \,w^2 -\frac{\lambda \sqrt{M\alpha }}{1+a\tau _3} \nonumber\\
B(w)&=& \frac{M}{1+a\tau _1}\,w^2-\frac{2\lambda \sqrt{M\alpha}}{1+a\tau _3}\,w+\frac{\alpha}{1+a\tau _2}.
\label{Bw}
\eea
Notice there is a relation between the functions $A(w), B(w)$ and $f(w)$ of the form 
\be
A(w)=f(w)+\frac{1}{2}\frac{d}{dw}B(w) .
\label{AB}
\ee
The stationary probability distribution (SPD) of the system can be obtained from Eqs.~\eqref{afpe}- \eqref{AB}) and can be written as \cite{gardiner}
\be
P_s(w)=\frac{N}{\sqrt{B(w)}}\, \exp\left[ \int^w{\frac{f(w')}{B(w')}dw'}\right] ,
\label{spd1}
\ee
where $N$ is the normalization constant that is determined by
\be
\int^\infty_0 P_s(w)\,dw=1 .
\label{norm}
\ee
Depending on the cross-correlation strength $\lambda$ one has to distinguish between different cases. 
The solution of the SPD for $0\leq \lambda <\frac{1+a\tau _3}{\sqrt{(1+a\tau _1)(1+a\tau _2)}}$ is
\bea
P_s(w)=\frac{N}{\sqrt{B(w)}}\, \exp\left[-\frac{\tilde{U}(w)}{M}\right] ,
\label{spd2}
\eea
where we have introduced a generalized potential according to 
\be
\tilde{U}(w)=\tilde{h}(w)-\tilde{E}\, \ln\left[ B(w)\right]-\frac{\tilde{F_1}-\tilde{F_2}}{\sqrt{\tilde{M}\tilde{\alpha }-\tilde{\lambda }^2}}\,\arctan\left[\frac{\tilde{M}w-\tilde{\lambda }}{\sqrt{\tilde{M}\tilde{\alpha }-\tilde{\lambda }^2}}\right] \,.
\label{spdU} 
\ee
Here, the following abbreviations are utilized:
\be
\tilde{M}=\frac{M}{1+a\tau _1},\, \tilde{\alpha}=\frac{\alpha }{1+a\tau _2},\, \tilde{\lambda}=\frac{\lambda \sqrt{M\alpha }}{1+a\tau _3}\,.
\ee
The non-universal exponent $\tilde{E}$ reads
\be
\tilde{E}=\frac{a(1+a\tau _1)}{2}-b_0\lambda \sqrt{\frac{\alpha}{M}}\, \frac{(1+a\tau _1)^2}{1+a\tau _3}-\frac{\beta M\lambda \sqrt{M\alpha}}{K(1+a\tau _3)}
\label{spdE}
\ee
with
\be
K=\left( \frac{M}{1+a\tau _1}-\frac{\alpha}{1+a\tau _2}\right)^2+\frac{4\lambda ^2M\alpha }{(1+a\tau _3)^2}\,.
\label{spdK}
\ee
Further we use
\bea
\tilde{F_1}&=& a\lambda \sqrt{M\alpha}\, \frac{1+a\tau _1}{1+a\tau _3}+b_0\alpha (1+a\tau _1)\, \left( \frac{1}{1+a\tau _2}-2\lambda ^2\, \frac{1+a\tau _1}{(1+a\tau _3)^2}\right) \nonumber\\
\tilde{F_2}&=&\frac{\beta M\alpha }{K}\left( \frac{\alpha }{(1+a\tau _2)^2}-\frac{M}{(1+a\tau _1)(1+a\tau _2)}+\frac{2\lambda ^2M\alpha }{(1+a\tau _3)^2}\right) \nonumber\\
\tilde{h}(w)&=&\frac{\beta M}{K}\left( \left( \frac{M}{1+a\tau _1}-\frac{\alpha }{1+a\tau _2}\right) \arctan w -\frac{\lambda \sqrt{M\alpha }}{1+a\tau _3}\, \ln\left[1+w^2\right] \right)
+\tilde{y}(w) \nonumber\\
\tilde{y}(w)&=& b_0(1+a\tau _1)\,w \,.
\label{spdg}
\eea
Let us remark, that by setting $\tau_1=\tau_2=0$ and $\beta=0$ one obtains the corrected solution for the SPD in \cite{mxz} (equations (19)-(22)).
In the present paper we assume that all correlation times take the same values, that is $\tau _1=\tau _2=\tau_3=\tau$ resulting in new expressions
for the generalized potential denoted now as $U(w)$. In case of the condition $0\leq \lambda <1$ we get  
\be
P_s(w)=\frac{N}{\sqrt{B(w)}}\, \exp\left[-\frac{U(w)}{M}\right] 
\label{spd3}
\ee
where B(w) according to Eq.~\eqref{Bw} changes to
\be
B(w)=\frac{M}{1+a\tau}\,w^2-\frac{2\lambda \sqrt{M\alpha}}{1+a\tau}\,w+\frac{\alpha}{1+a\tau}.
\label{Bw2}
\ee
The potential takes the form
\be
U(w)=h(w)-E\, \ln\left[ B(w)\right]-\frac{F(1+a\tau )}{\sqrt{M\alpha (1-\lambda ^2)}}\,\arctan\left[\frac{M\, w-\lambda \sqrt{M\alpha }}{\sqrt{M\alpha (1-\lambda ^2)}}\right] 
\label{spdU2} 
\ee
with 
\bea
E &=& \frac{a(1+a\tau)}{2}-\left( b_0+\frac{\beta M^2}{Q}\right)\, \lambda \sqrt{\frac{\alpha}{M}}\,(1+a\tau ) \nonumber\\
Q &=& M^2+\alpha ^2+2M\alpha (2\lambda ^2-1) \nonumber\\
F&=& a\lambda \sqrt{M\alpha}-b_0\alpha \left( 2\lambda ^2-1\right) -\frac{\beta M\alpha \left( \alpha +M(2\lambda ^2-1)\right)}{Q} \nonumber\\
h(w)&=& \frac{\beta M(1+a\tau )}{Q} \left( (M-\alpha ) \arctan w-\lambda \sqrt{M\alpha }\, \ln\left[1+w^2\right] \right) + y(w) \nonumber\\
y(w)&=& b_0(1+a\tau )\,w .
\label{spdg2}
\eea
We want to point out that setting $\beta=0$ and $\tau=0$ gives the right results for the SPD that is not correct in \cite{qw}. These solutions are in agreement with those mentioned in \cite{bfr}.
The second case of $\lambda =1$ has to be considered separately. The corresponding solution is
\be
P_s(w)=\frac{N}{\sqrt{B(w)}}\,\exp\left[-\frac{U^*(w)}{M}\right]
\label{spd1b}
\ee
and the generalized potential reads 
\be
U^*(w)=h^*(w)-E^*\, \ln\left[ B(w)\right] -\frac{F^*(1+a\tau)}{\sqrt{M\alpha }-M\, w}\,. 
\label{Ustern}
\ee
The non-universal exponent is written in the form
\be
E^*=\frac{a(1+a\tau)}{2}-\left( b_0+\frac{\beta M^2}{Q}\right)\, \sqrt{\frac{\alpha}{M}}\,(1+a\tau )
\label{Estern}
\ee
with
\bea
Q^* &=& (M+\alpha )^2 \nonumber\\
F^* &=& a\sqrt{M\alpha }-b_0\alpha -\frac{\beta M\alpha (\alpha +M)}{Q^*}\nonumber\\
h^*(w)&=& \frac{\beta M(1+a\tau )}{Q^*} \left( (M-\alpha ) \arctan w -\sqrt{M\alpha }\, \ln\left[1+w^2\right] \right) + y(w) \,.
\label{hstern}
\eea
The function $y(w)$ remains unchanged and is given by Eq.~\eqref{spdg2}.
In the following sections we only consider the first case and analyze the results for $0\leq \lambda <1$, i.e. our further computations refer 
to the solutions given by Eqs.~(\ref{spd3}) - (\ref{spdg2}).

\subsection{Properties of the SPD} 

In this section we discuss the behavior of the stationary probability distribution (SPD) calculated analytically in the previous subsection. 
In Fig.~\ref{SPDBeta} the SPD is represented as function of the tumor cell population $w$ under different immunization rates $\beta$. 
The SPD reveals a maximum indicating the most probable cell population. The maximum becomes the more pronounced the higher the immunization rate 
$\beta$ is. The maximum is shifted to smaller tumor population with increasing rate $\beta$.  
\bef[ht]
	\includegraphics[width=10cm]{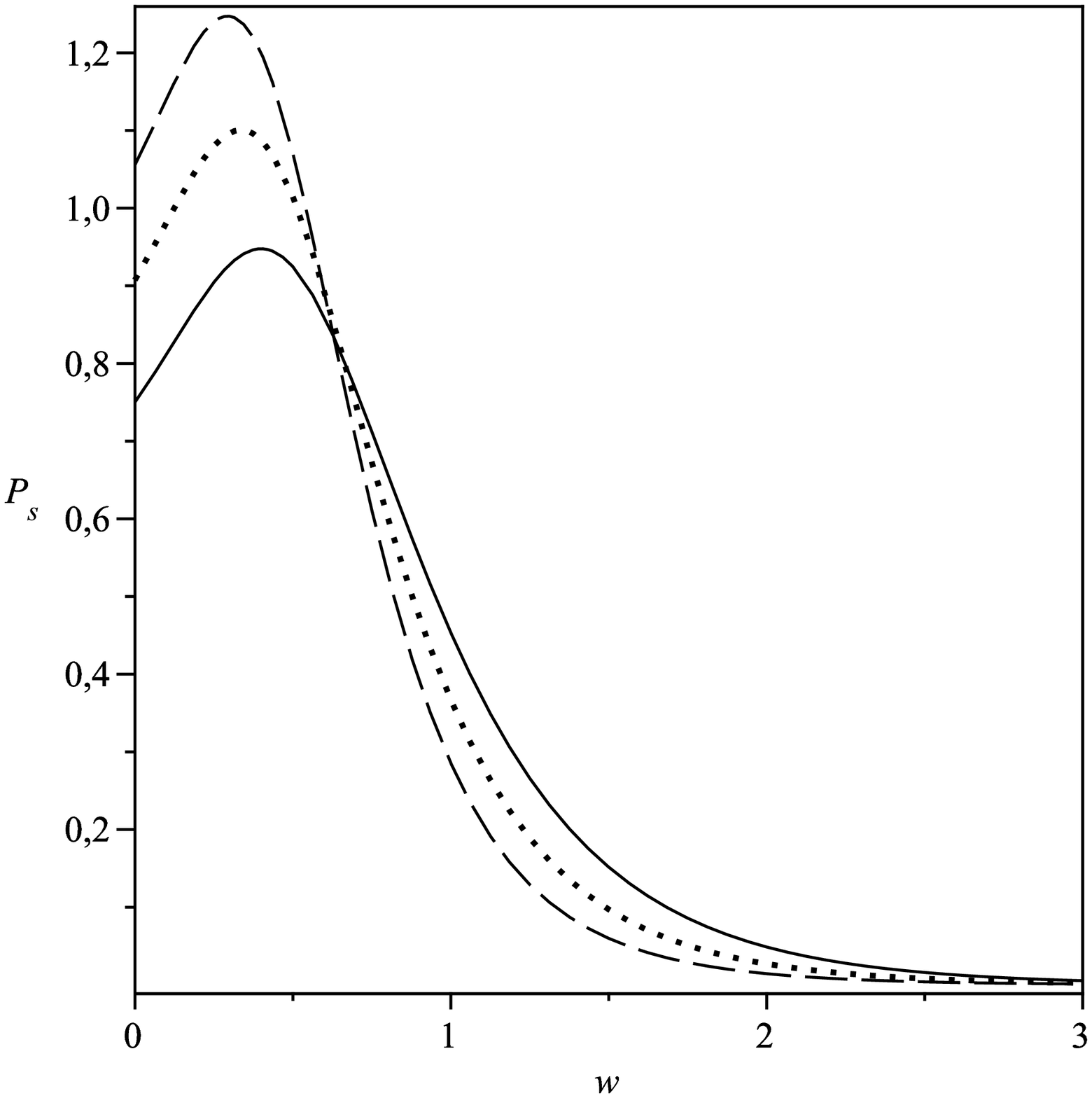}
	\caption{Plot of the SPD $P_s$ as a function of the cell population $w$ for fixed $a=0.5$, $b_0=1.0$, $\alpha=0.3$, $M=0.7$, $\tau=0.5$ and 
$\lambda=0.5$. The immunization $\beta$ varies from $0.0$ (solid line), $0.5$ (dotted line), $1.0$ (dashed line).}
	\label{SPDBeta}
\ef
The SPD is influenced significantly by the cross-correlation characterized by the parameter $\lambda$. The maximum is strongly enhanced by an increasing 
cross-correlation strength as shown in Fig.~\ref{SPDLambda}.  
\bef[ht]
	\includegraphics[width=10cm]{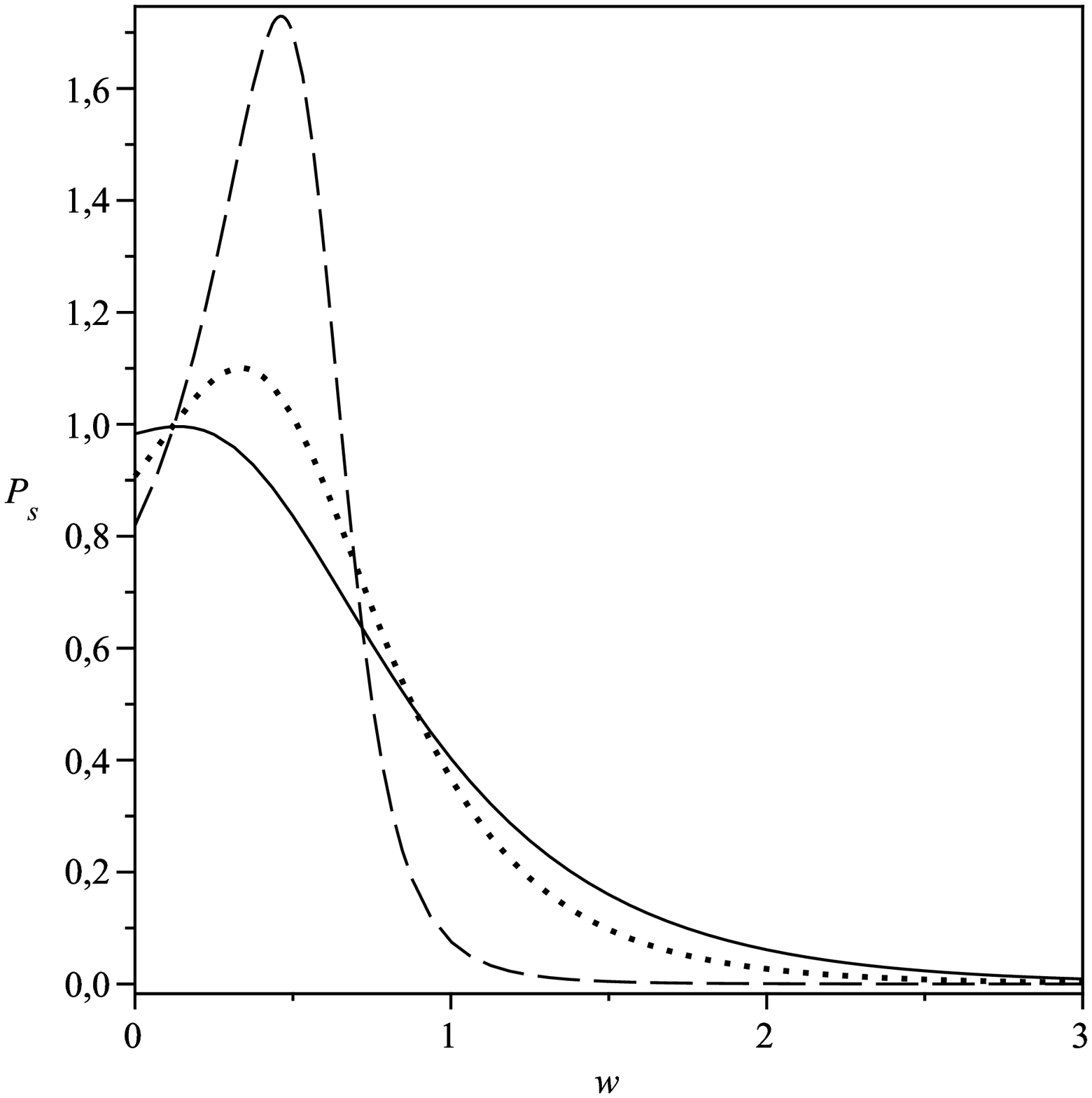}
	\caption{Plot of the SPD $P_s$ as a function of the cell population $w$ for fixed $a=0.5$, $b_0=1.0$, $\alpha=0.3$, $M=0.7$, $\tau=0.5$ and $\beta=0.5$.   The strength of the cross-correlation $\lambda$ takes $0.1$ (solid line), $0.5$ (dotted line), $0.9$ (dashed line).}
	\label{SPDLambda}
\ef
The SPD is also influenced by the correlation time $\tau$ of the noises. The result is shown in Fig.~\ref{SPDTau}.
\bef[ht]
	\includegraphics[width=10cm]{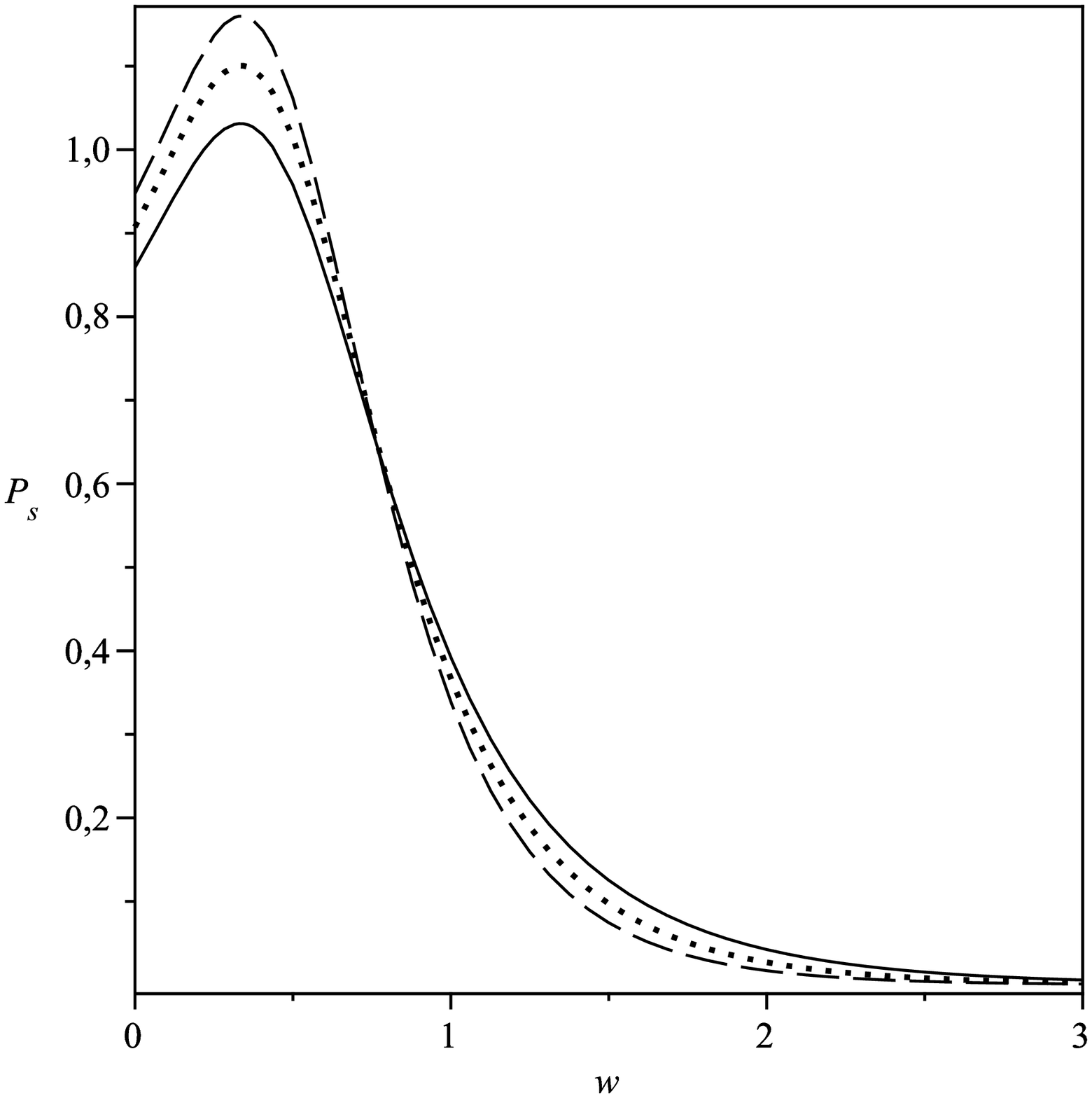}
	\caption{Plot of the SPD $P_s$ as a function of the cell number $w$ when $a=0.5$, $b_0=1.0$, $\lambda=0.5$, $M=0.7$, $\alpha=0.3$ and $\beta=0.5$ 
are fixed: $\tau$ takes $0.0$ (solid line), $0.5$ (dotted line), $1.0$ (dashed line).}
	\label{SPDTau}
\ef
There appears already a maximum which is more articulated when the correlation time is enhanced.

\subsection{Biological interpretation}

The importance of an efficient immunization against tumor evolution is illustrated in Fig.~\ref{SPDBeta}. This efficacy depends on the competence 
of the immune system to detect the malignant cancer cells, and thus to initiate a power full immune response. 
In \cite{dunn1,dunn2} the 'Three E's of cancer immunoediting' are described, i.e. the tumor-immune interaction can result into three different 
phases: elimination, equilibrium and escape, whereas sooner or later the equilibrium phase offers a cross-over to the other phases.
Adopting this concept to the behavior of the SPD the tumor elimination phase is the more probable and the escape phase is the more improbable 
the higher the immune coefficient $\beta$ is, for further remarks compare also subsection IV.C.

Now, we want to relate the internal noise $\epsilon_1$ and the external noise $\epsilon_2$, introduced in Eq.~\eqref{lg}, to real processes that 
occur in the tumor and its environment including the hallmarks of cancer \cite{hawei}, and moreover to point out to the connections among each other. 
External noise is thought to be originated in the extracellular matrix embedding the tumor or it is a consequences of drug delivery from outside 
of the host. Additionally, external noise can be caused by thermal fluctuations. In contrast the internal noise is generated directly within the 
tumor system as a kind of self-organization, for instance by gene mutations resulting in a multitude of genetically different tumor cells within 
the same system. The process is based 
upon internal mechanisms inside the tumor without contact to its environment. Although the origins of both stochastic processes are different  
one should argue that there exists an interrelation among both ones. A measure for such a correlation is the strength of the cross-correlation denoted by 
$\lambda$ in Eq.~\eqref{lg1} as well as the correlation time $\tau$. The expected coupling between external and internal noises can be understood as 
follows. The normal tissue adjacent to the malignant one produces anti-growth signals in order to avoid an uncontrolled growth. The tumor may 
respond with insensitivity with respect to these signals by alteration or down-regulation of the corresponding receptors. Furthermore, 
some tumor cells are able to develop self-sufficiency in generating growth signals. Another correlation concerns the nutrient supply. With a 
growing tumor tissue the competition is intensified regarding the nutrients between normal tissue and the nascent transformed cells. The tumor 
can sustain and induce angiogenesis via an 'angiogenic switch' from vascular quiescence in order to progress to a larger size. Another characteristic 
of tumor growth is the acquisition of a diversity of strategies to evade apoptotic signals that are emitted on the one hand by the tumor environment 
and on the other hand generated within the tumor cells. 

Therefore, the behavior of the SPD depending on the strength of the cross-correlation is clearly shown in Fig.~\ref{SPDLambda}. An increasing 
$\lambda$ is equated with an increasing ability of the tumor to compensate the external interferences via internal reactions described above. Thus, 
in case of strong correlations the tumor has an improved ability to reach the escape phase. 

In order to explain the dependence of our results on the correlation time let us remind that $\tau$ is the correlation time of the 
cross-correlation as well as the correlation time of the auto-correlation functions of the additive (external) and multiplicative (internal)
noise, respectively. Here we have assumed that the correlation time for both kind of noises is relevant on the same time scale $\tau$. 
Taking this into account the appearance of a finite correlation time leads to a higher probability of a certain tumor size but does not change 
the likeliest tumor size as presented in Fig.~\ref{SPDTau}.

Notice that we attribute a random nature to the mechanisms of the tumor evolution because the details of the growth and decay processes 
differs from patient to patient. Therefore, tumor growth and the interplay with the environment can be regarded as a stochastic process 
and is interpreted by introducing external and internal noises.

\section{Mean-First-Passage Time (MFPT)}

\subsection{Derivation of the MFPT}

In cancer treatment it is of interest whether a tumor that reached a certain size can suffers extinction by external or internal interferences, 
i.e. is it possible that the influences of the noises and the immune system introduced in the previous sections can cause extinction of the tumor. A further concern is the transition time between these two states: the lethal tumor size and the tumor free state, respectively. In order to describe these transient properties of the system we apply the mean-first-passage time that is given by the following expression \cite{liwe,maliwe}
\be
T_{w_1 w_2}= \int\limits_{w_1}^{w_2}{\frac{dw}{B(w)P_s(w)}}\int\limits_{w}^{\infty}{P_s(v)dv},
\label{mfpt1}
\ee
i.e. the transition from an initial point $w_1$ to an end point $w_2$ is considered. We choose the stationary points of the effective potential (\ref{Vsol}), more specifically $w_1= x_s$ and $w_2=0$, i.e. the MFPT of the system reaching the tumor free state is studied. Now we make use of an approximation scheme that is valid for small $M$ and $\alpha$ in comparison with the potential barrier high $\left[ U(w_2)-U(w_1 )\right]$ \cite{gardiner,gumi} what has already been applied to tumor models, e.g. \cite{wwm}. We derive an analytical expression for (\ref{mfpt1}), namely
\be
T_{w_1 w_2}= \frac{2\pi}{\sqrt{\mid V''(0)V''(w_1)\mid }}\, \exp\left[ \frac{1}{M}\left[ U(0)-U(w_1)\right] \right]\,,
\label{mfpt2}
\ee 
where the double-prime denote the second derivation with respect to $w$. Inserting Eq.~\eqref{Vsol} and Eqs.~\eqref{spdU2}-\eqref{spdg2} 
into the Eq.~\eqref{mfpt2} leads to the final expression
\bea
T_{w_{1} w_{2}}&=& \frac{2\pi }{a R}\,
\exp\left\{ \frac{1}{M} \left[ E\, \ln \frac{B(w_1)}{B(0)} - h(w_1) \right . \right . \nonumber\\
&+& \left . \left . \frac{F(1+a\tau )}{\sqrt{M\alpha (1-\lambda ^2)}}\, 
\left( \arctan \frac{\lambda \sqrt{M\alpha }}{\sqrt{M\alpha (1-\lambda ^2 ) }} 
- \arctan \frac{\lambda \sqrt{M\alpha }-M w_1}{\sqrt{M\alpha (1-\lambda ^2)}} \right) \right]  \right\}
\label{mfpt3}
\eea
where
\bea
R&=&\sqrt{\frac{2 w_1}{a}\left(  b_0+\frac{\beta }{\left( 1+{w_1}^2\right) ^2}\right) -1}\nonumber\\
B(w_1)&=&\frac{M}{1+a\tau}\,{w_1}^2-\frac{2\lambda \sqrt{M\alpha}}{1+a\tau}\,w_1+\frac{\alpha}{1+a\tau},\quad B(0)=\frac{\alpha}{1+a\tau} \nonumber\\
h(w_1)&=&\frac{\beta M(1+a\tau )}{Q} \left( (M-\alpha ) \arctan w_1 - \lambda \sqrt{M\alpha }\, \ln\left[1+{w_1}^2\right] \right) + y(w_1)\nonumber\\ 
y(w_1)&=& b_0 w_1 (1+a\tau ).
\eea
Both constants, $E$ and $F$, are still the same as those in Eq.~\eqref{spdg2}. 
Notice, that applying our solutions obtained by Eqs.~\eqref{spdU} - \eqref{spdg} into Eq.~\eqref{mfpt2}, therefore substituting $U(0)$ and $U(w_1)$ by $\tilde{U}(0)$ and $\tilde{U}(w_1)$, respectively, and setting $\tau _1=\tau _2=0$ and $\beta=0$ yields the correction of the expression in Eq.~(27) 
in \cite{mxz}.

\subsection{Properties of the MFPT}

In this subsection we discuss the properties of our system and the behavior of the MFPT. In Fig.~\ref{TvsM} we present the MFPT as function of the parameter 
$M$ introduced in Eq.~\eqref{lg1}. This parameter $M$ is a measure for both the auto- and the cross-correlation function between internal and external 
noise. 
\bef[ht]
	\includegraphics[width=10cm]{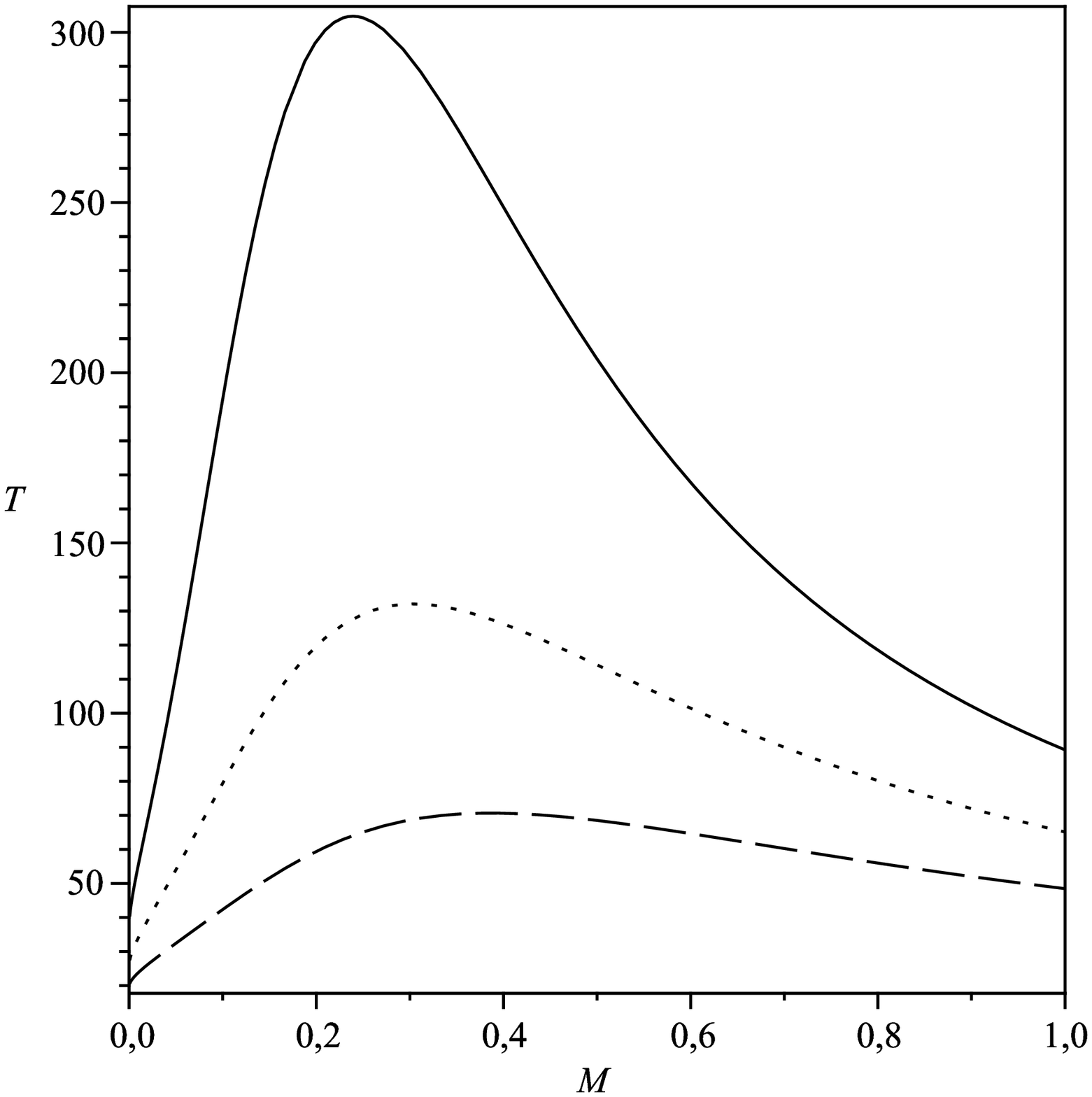}
	\caption{Plot of the MFPT as a function of $M$ when for fixed values $a=1.0$, $b_0=1.0$, $\tau=0.5$, $\alpha=0.1$ and $\lambda=0.8$. 
The immunization strength $\beta$ varies from $0.3$ (solid line), $0.5$ (dotted line), $0.7$ (dashed line).}
	\label{TvsM}
\ef
As a feature there occurs a maximum indicating a long living cell population. The maximum is the more pronounced the lower the immunization rate is 
and it is shifted towards higher values of $M$. Increasing the rate $\beta$ the MFTP is smaller and an extinction of the tumor population is more probable. 
In Fig.~\ref{TvsAlpha} the MFPT is represented depending on the parameter $\alpha$ according to Eq.~\eqref{lg1}. Here $\alpha$  
characterizes the strength of the auto-correlation of the additive noise as well as the strength of cross-correlation. The increase of 
$\alpha $ leads to a decrease of the MFPT. This decay is very strong in case of a high immunization rate as expected. 
\bef
	\includegraphics[width=10cm]{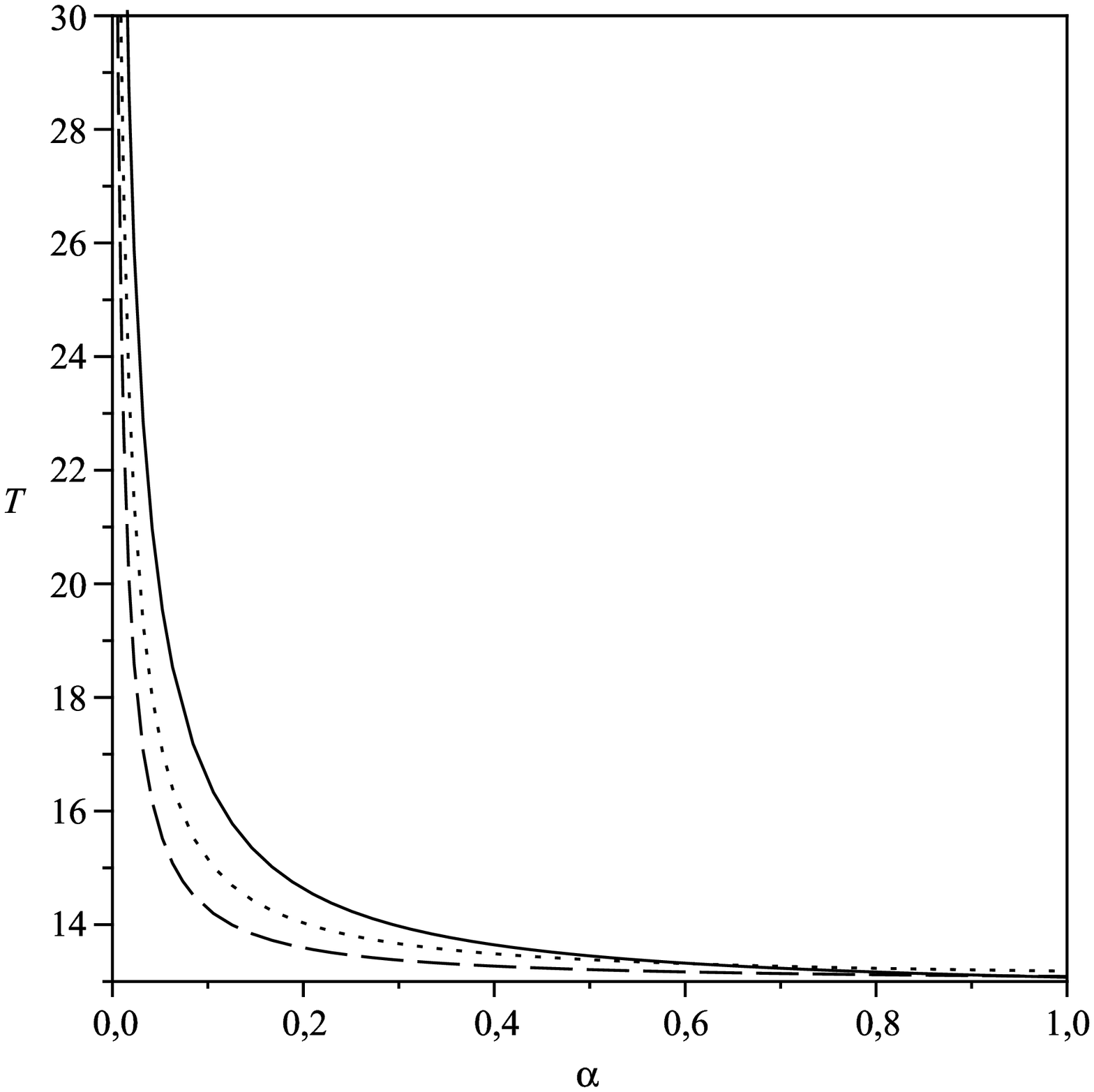}
	\caption{Plot of the MFPT as a function of $\alpha $ when $a=0.5$, $b_0=1.0$, $\tau=0.5$, $M=0.8$ and $\lambda=0.5$ are fixed. The parameter  
$\beta$ takes $0.1$ (solid line), $0.5$ (dotted line), $0.9$ (dashed line).}
	\label{TvsAlpha}
\ef
The direct influence of the immunization strength $\beta$ on the MFPT is shown in Fig.~\ref{TvsBetaTau}. There appears already a maximum which is shifted 
to higher values of $\beta$ when the correlation time $\tau$ is reduced. A similar behavior of the MFPT as function of $\beta$ is also observed in 
dependence on the parameter $\lambda$. 
\bef
	\includegraphics[width=10cm]{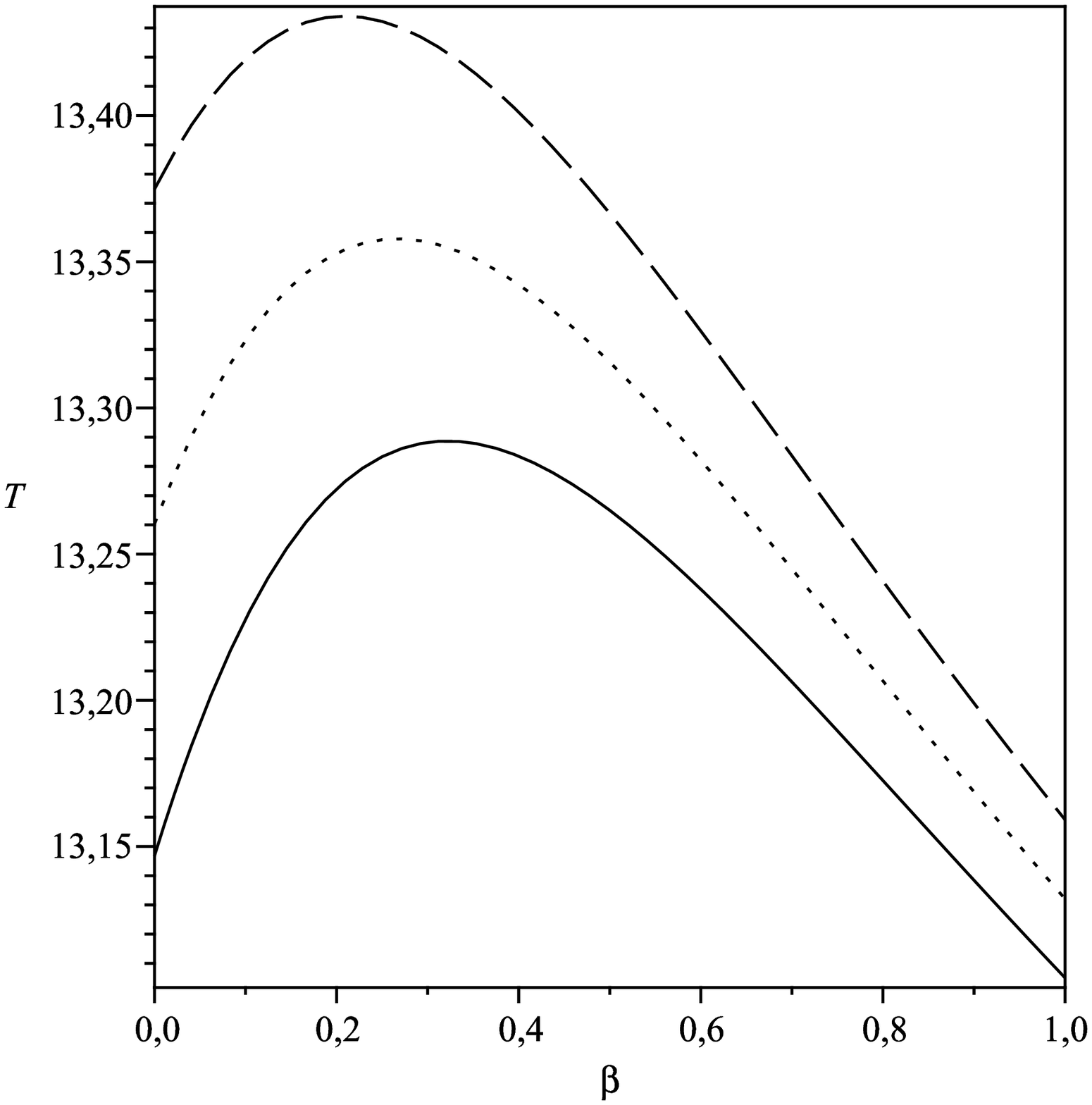}
	\caption{Plot of the MFPT as a function of $\beta $ for fixed values for $a=0.5$, $b_0=1.0$, $\alpha=0.6$, $M=0.8$ and $\lambda=0.5$. 
The correlation time $\tau$ varies from $0.1$ (solid line), $0.5$ (dotted line), $0.9$ (dashed line).}
	\label{TvsBetaTau}
\ef
A very instructive behavior can be observed in Fig.~\ref{TvsBetaM} where the MFPT is depicted as function of the immunization coupling $\beta$ with 
variation of the global noise strength $M$. The maximum becomes more pronounced if the noise strength increases. 
\bef
	\includegraphics[width=10cm]{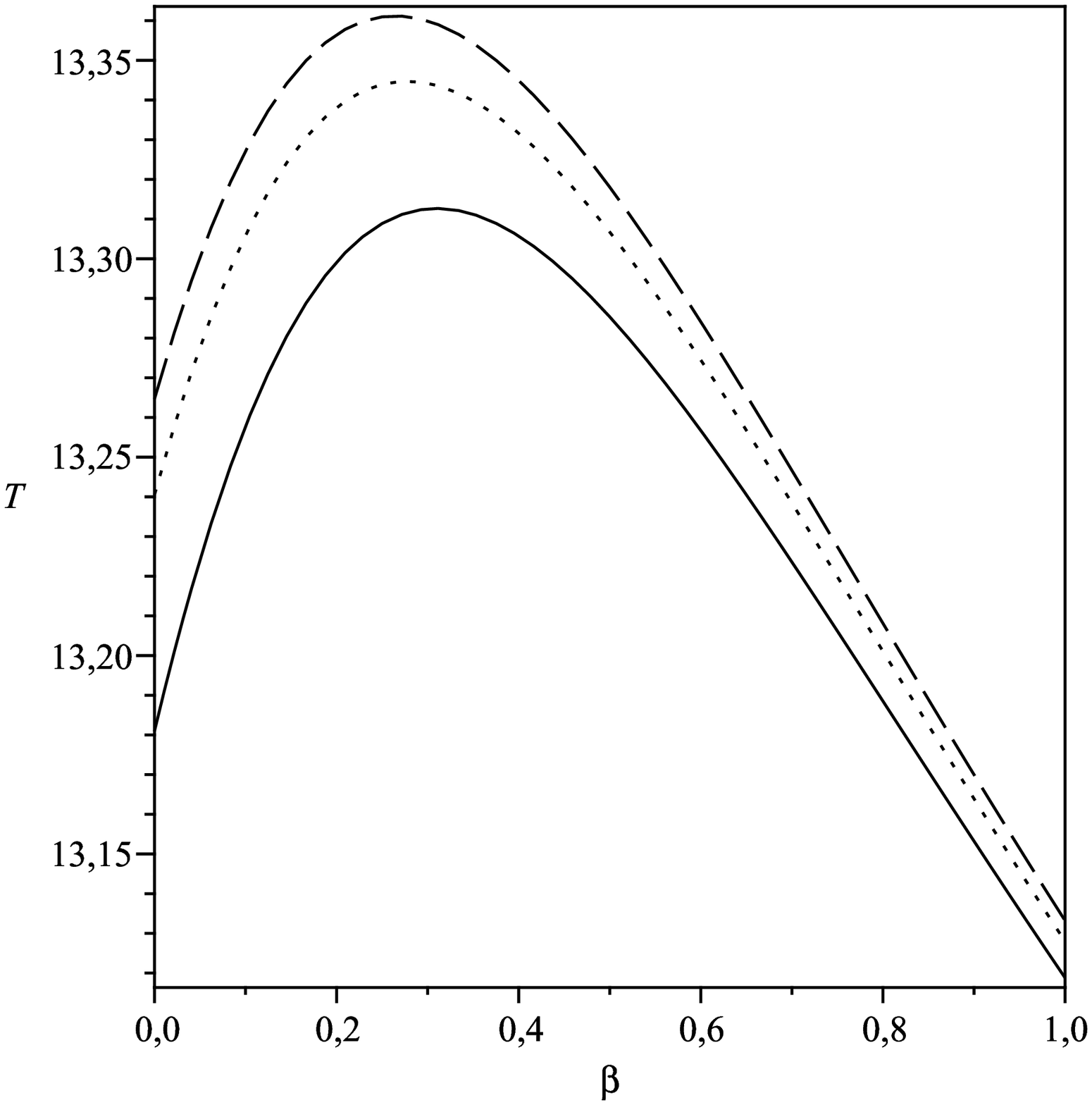}
	\caption{Plot of the MFPT as a function of $\beta $ when $a=0.5$, $b_0=1.0$, $\alpha=0.6$, $\lambda=0.5$ and $\tau=0.5$ are fixed. The noise strength 
$M$ is $0.1$ (solid line), $0.5$ (dotted line), $0.9$ (dashed line).}
	\label{TvsBetaM}
\ef
A nearly linear behavior of the MFPT as function of the correlation time $\tau$ is observed in Fig.~\ref{TvsTauBeta}. The increase of the MFTP is weaker 
for a stronger immunization rate $\beta$. 
\bef
	\includegraphics[width=10cm]{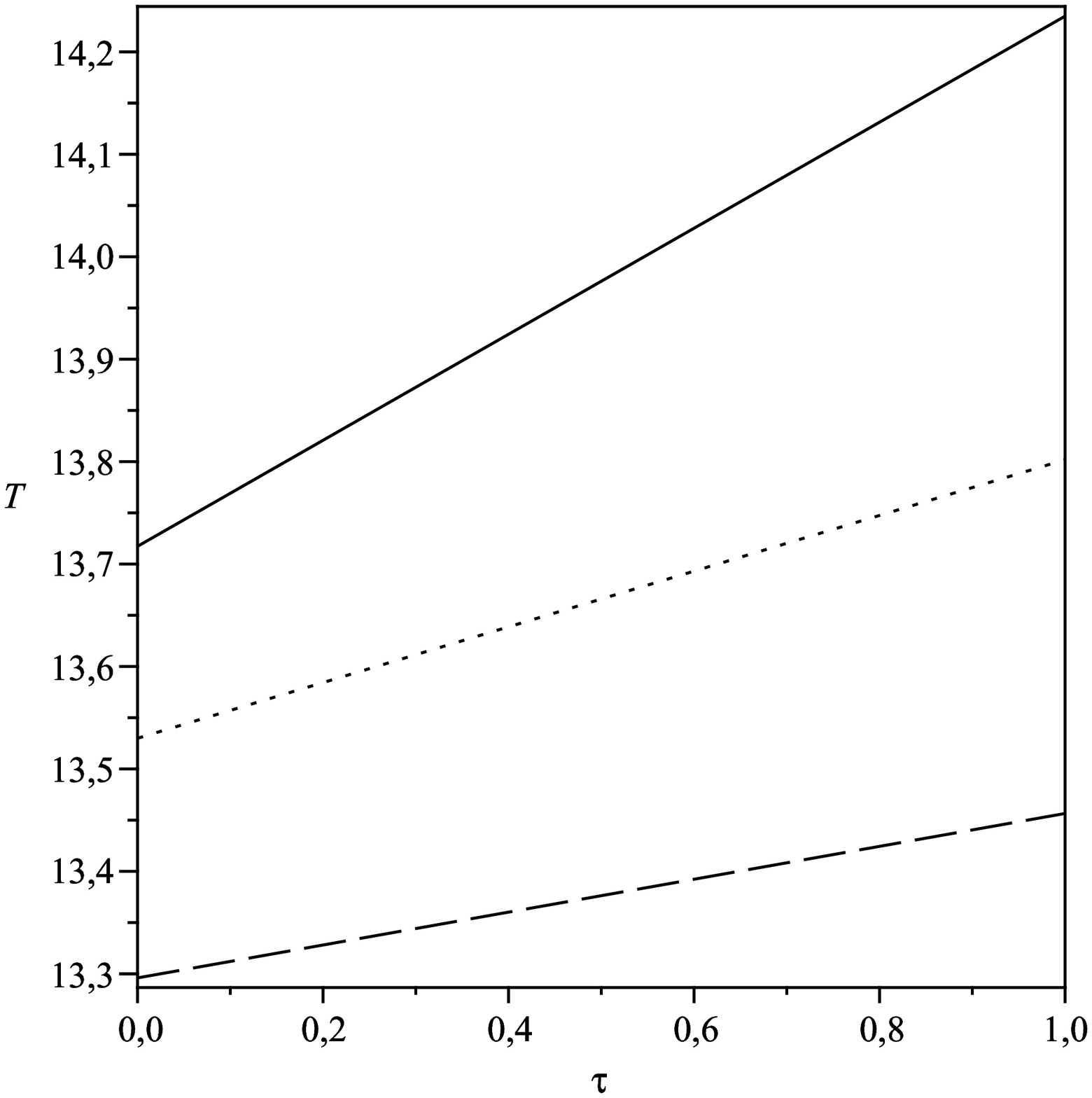}
	\caption{Plot of the MFPT as a function of $\tau$ when $a=0.5$, $b_0=1.0$, $\alpha=0.3$, $M=0.8$ and $\lambda=0.5$ are fixed: $\beta$ takes $0.1$ (solid line), $0.5$ (dotted line), $0.9$ (dashed line).}
	\label{TvsTauBeta}
\ef

\subsection{Biological aspects}

In this subsection the behavior of the MFPT is discussed with regard to biological aspects. Let us stress that a decrease of the MFPT is 
tantamount to an increase of the probability of the transition to the tumor free state. At first, we consider the influence of the 
multiplicative noise on the MFPT and its relation to the immune system, where the noise is originated from all the distinct processes  
described in subsection III.C. Here, we assume that the multiplicative noise is mainly determined by the gene mutations. Fig.~\ref{TvsM} 
indicates that there exists an appropriate $M$ leading to a high MFPT. Improving the effectiveness of the immune system leads to a reduction 
of the MFPT. Moreover, the tumor system requires more gene mutations in order to maximize the MFPT. More genetic alterations induce a 
deterioration of the ability of the immune system to identify tumor cells. But this mechanism is limited as it is visible by the descent of the 
curves in Fig.~\ref{TvsBetaTau}. As soon as the optimal value of the strength of the multiplicative noise is exceeded the MFPT decreases and 
consequently the ability of the self-organization is reduced.

The influence of the external (additive) noise offers the expected behavior. All interferences impair the living conditions of the tumor. Therefore, 
a growing parameter $\alpha$ leads to a decline of the MFPT and enhances the probability of the extinction of the cancer.

Following the explanation made for the interpretation of Figs.~\ref{TvsBetaTau} and ~\ref{TvsBetaM} let us introduce the principle of 
immunoediting presented in \cite{dunn1,dunn2}. On the one hand the immune system is able to cause extinction of the tumor, otherwise it can 
facilitate tumor progression by sculpting the immunogenic phenotype of tumors. During this process immune-resistent variants of the tumor 
cells are able to survive and even more proliferate in order to develop a tumor tissue that can sustain further immune attacks. This behavior is displayed 
in Figs.~\ref{TvsBetaTau} and ~\ref{TvsBetaM}. The increasing part of the curve is thought to be connected with the process of tumor 
sculpting which may end up in the tumor escape phase. The decreasing part of the MFPT is identified with immunosurveillance that leads 
to tumor elimination. This process is the more likely the bigger the immunization strength is. The effect of the strength of the 
multiplicative noise $M$ and the strength of the cross-correlation $\tau$ are similar. Due to the fact that the increase of both 
parameters $M$ and $\tau$ and the biological mechanisms beyond promotes the tumor growth, one should expect a retardation of the transition 
to the elimination phase as shown in Figs.~\ref{TvsBetaTau} and ~\ref{TvsBetaM}. 

The correlation-time $\tau$ also effects the MFPT. An increase of $\tau$ leads to a slowing down of the transition between the different 
states of the tumor. The longer the correlation time $\tau$ is the more probable are long living tumor populations. Consequently, a rising 
value of $\tau$ simplifies the opportunity of the tumor to evade the immune system.

\section{Conclusions}

\noindent In this work we have proposed and analyzed a more refined model describing tumor cell growth. Starting from a logistic 
model we have modified the model in several directions. The decay term is supplemented by a deterministic non-linear immunization term which 
enhances the death rate of the tumor. Furthermore, the birth rate as assumed to be stochastically distributed leading to a multiplicative noise. 
The occurrence of such a noise term is motivated by the underlying biological situation. Additionally, the system is subjected to an additive, 
external noise which is originated by the external conditions as the environment of the tumor. Both kinds of colored noises are correlated, i.e. 
there are autocorrelation and a cross-correlation functions with different strength. The resulting equation has the form of a 
Langevin equation which can be transformed into a Fokker-Planck equation. Using standard methods we find the steady state solutions 
which are discussed depending on the strength of the cross-correlation, the finite correlation time and the degree of immunization. 
The behavior of the stationary probability distribution is analyzed taking into account biological aspects above all the three different 
states of the tumor: elimination, equilibration and escape phase. In particular, the SPD offers a maximum 
indicating the appearance of very probable states. This maximum becomes for instance the more pronounced the higher the immunization rate is. 
As a further quantity of interest we have studied the mean-first passage time which indicates when the tumor suffers extinction. The MFPT is 
likewise calculated analytically and analyzed under consideration of biological aspects. The MFPT is influenced in a significant manner by the 
immunization and the cross-correlation as well as the finite correlation time of the underlying colored noises. The observed behavior can be 
understood in terms of the above mentioned three phases of a tumor population.

\begin{acknowledgments}
We are grateful to Prof.D.Vordermark and Dr.F.Erdmann and for valuable discussions and experimental realizations.
\end{acknowledgments}


\clearpage
\newpage
\begin{thebibliography}{90}

\bibitem{murray} J.D. Murray, {\em Mathematical Biology}. Springer Verlag, Berlin, 1993\,.

\bibitem{lo} A.J. Lotka, J.Am.Chem.Soc {\bf 42}, 1595 (1920)\,.

\bibitem{vo} V. Volterra, Atti R. Accad.Naz.Lincei, Mem.Cl.Sci.Fis.Mat.Nat.{\bf 2}, 31 (1926)\,.

\bibitem{cl} G.Q.Cai, and Y.K.Lin, Phys.Rev. E {\bf 70}, 041910 (2004)\,.

\bibitem{rmf} T.Reichenbach, M.Mobilia, and E.Frey, Phys.Rev. E {\bf 74}, 051907 (2006)\,.

\bibitem{plg} S.Pigolotti, C.L\'opez, and E.Hern\'andez-Garc\'ia, Phys.Rev.Lett. {\bf 98}, 258101 (2007)\,.

\bibitem{mgt} M.Mobilia, I.T.Georgiev, and U.C.T\"auber, Phys.Rev. E {\bf 73}, 040903(R) (2006)\,.

\bibitem{as} R.Abta and N.M.Shnerb, Phys.Rev. E {\bf 75}, 051914 (2007)\,.

\bibitem{rs} P.A.Rikvold and V.Sevim, Phys.Rev. E {\bf 75}, 051920 (2007)\,.

\bibitem{bs} N.Brenner and Y.Shokef, Phys.Rev.Lett. {\bf 99}, 138102 (2007)\,.

\bibitem{mbf} M.Maru\v{s}i\'c, \v{Z}.Bajzer, S.Vur-Palovi\'c, and J.P.Feyer, Bull. Mathematical Biology {\bf 56}, 617 (1994)\,.

\bibitem{mxz} D.C.Mei, C.W.Xie, and L.Zhang, Eur.Phys. J. B {\bf 41}, 107 (2004)\,.

\bibitem{qw} Bao-Quan Ai, Xian-Ju Wang, Guo-Tao Liu, and Liang-Gang Liu, 
Phys.Rev E {\bf 67}, 022903 (2003)\,.

\bibitem{bfr} A.Behera and S.F.O'Rourke, Phys.Rev E {\bf 77}, 013901 (2008)\,.

\bibitem{qw2} Bao-Quan Ai, Xian-Ju Wang and Liang-Gang Liu, 
Phys.Rev E {\bf 77}, 013902 (2008)\,.

\bibitem{zsh} Wei-Rong Zhong, Yuan-Zhi Shao, and Zhen-Hui He, Phys.Rev. E {\bf 73}, 060902(R) (2006)\,.

\bibitem{wwm} Can-Jun Wang, Qun Wei, and Dong-Cheng Mei, Modern Physics Letters B {\bf 21}, 789 (2007)\,.

\bibitem{kbbd} F.Kozusko, M.Bourdeau, Z.Bajzer, and D.Dingli, Bull. Mathematical Biology {\bf 69}, 1691 (2007)\,.

\bibitem{zslwh} Wei-Rong Zhong, Yuan-Zhi Shao, Li Li, Feng-Hua Wang, and Zhen-Hui He, Euro.Phys.Lett. {\bf 82} ,20003 (2008)\,.

\bibitem{gardiner} C.W. Gardiner, {\em Handbook of Stochastic Methods}. Springer Verlag, Berlin, 1990\,.

\bibitem{wlk} Da-jin Wu, Li Cao, and Sheng-zhi Ke, Phys.Rev. E {\bf 50}, 2496 (1994)\,.

\bibitem{jl} Ya Jia and Jia-rong Li, Phys.Rev. E {\bf 53}, 5786 (1996)\,.

\bibitem{z} Ping Zhu, Eur.Phys. J. B {\bf 55}, 447 (2007)\,.

\bibitem{cb} H.Calisto and M.Bologna, Phys.Rev. E {\bf 75}, 050103(R) (2007)\,.

\bibitem{nm} L.R.Nie and D.C.Mei, Euro.Phys.Lett {\bf 79}, 20005 (2007)\,.

\bibitem{cdg} P.Castorina, P.P.Delsanto, and C.Guiot, Phys.Rev.Lett. {\bf 96}, 188701 (2006)\,.

\bibitem{am} M.Assaf and B.Meerson, Phys.Rev.Lett. {\bf 97}, 200602 (2006)\,.

\bibitem{bdsp} I.Bena, M.Droz, J.Szwabi\'nski, and A.P\c{e}kalski, Phys.Rev. E {\bf 76}, 011908 (2007)\,.

\bibitem{jms} D.E.Juanico,C.Monterola, and C.Saloma, Phys.Rev. E {\bf 75}, 045105(R) (2007)\,.

\bibitem{vk} N. G. van Kampen, {\em Stochastic Processes in Physics and Chemistry} (North-Holland, 
Amsterdam, 1992). 

\bibitem{dunn1} G.P.Dunn, A.T.Bruce, H.Ikeda, L.J.Old and R.D.Schreiber, Nat.Immunol. {\bf 3}, 991 (2002)\,.

\bibitem{dunn2} G.P.Dunn, L.J.Old and R.D.Schreiber, Annu.Rev.Immunol. {\bf 22}, 329 (2004)\,.

\bibitem{hawei} D.Hanahan and R.A.Weinberg, Cell {\bf 100}, 57 (2000)\,.

\bibitem{liwe} K.Lindenberg and B.J.West, J.Stat.Phys. {\bf 42}, 201 (1986)\,.

\bibitem{maliwe} J.Masoliver, B.J.West, and K.Lindenberg, Phys.Rev. A {\bf 35}, 3086 (1987)\,.

\bibitem{gumi} E.Guardia and M.S.Miguel, Phys. Lett. {\bf 109A}, 9 (1985)\,.


\end{thebibliography}
\end{document}